\documentclass[12pt]{article}
\usepackage[dvips]{graphicx,color}
\usepackage{amsmath}
\usepackage{amsfonts}
\begin{document}
\title{Planar field theories with space-dependent noncommutativity}
\author{C.~D.~Fosco and  G.~Torroba  \\
  {\normalsize\it Centro At\'omico Bariloche and Instituto Balseiro}\\
  {\normalsize\it Comisi\'on Nacional de Energ\'\i a At\'omica}\\
  {\normalsize\it 8400 Bariloche, Argentina.}}
\date{\today}
\maketitle
\begin{abstract}
\noindent
We study planar noncommutative theories such that the spatial coordinates
${\hat x}_1$, ${\hat x}_2$ verify a commutation relation of the form: $[{\hat
  x}_1, {\hat x}_2] = i \, \theta ({\hat x}_1,{\hat x}_2)$.  Starting
from the operatorial representation for dynamical variables in the
algebra generated by ${\hat x}_1$ and ${\hat x}_2$ , we introduce a
noncommutative product of functions corresponding to a specific
operator-ordering prescription.  We define derivatives and traces, and
use them to construct scalar-field actions.  The resulting expressions
allow one to consider situations where an expansion in powers of $\theta$
and its derivatives is not necessarily valid. In particular, we study
in detail the case when $\theta$ vanishes along a linear region. We show
that, in that case, a scalar field action generates a boundary term,
localized around the line where $\theta$ vanishes.
\end{abstract}
\bigskip
\newpage
\section{Introduction}\label{intro}
Noncommutative Quantum Field Theories~\footnote{See, for
example~\cite{DN,szabo} for pedagogical reviews.} have recently
attracted renewed attention, not only because of their relevance to
String Theory~\cite{DN,castellani}, but also in the Condensed Matter
Physics context, since they have been proposed as effective
descriptions of the Laughlin states in the Quantum Hall
Effect~\cite{susskind,PP,KS}.  Noncommutativity has also been
introduced to describe the skyrmionic excitations of the Quantum Hall
ferromagnet at $\nu=1$~\cite{PA, LMR}.

A planar system of charged particles in the presence of an external
magnetic field has a very rich structure, in part because of the
peculiarities of the Landau level spectrum for a single
particle~\cite{landau}. A noncommutative description is usually
invoked as a way to describe a restriction to the lowest Landau level,
a step which is justified by the existence of a large gap between the
lowest and higher Landau Levels~\cite{gj,flc}. This restriction cannot
be introduced as a smooth limit of the full (all level) system, since
there is a change in the number of physical degrees of freedom, an
effect that has been known since the early studies on Chern-Simons
Quantum Mechanics~\cite{jackiw}, and which is entirely analogous to
the reduction from the Maxwell-Chern-Simons action into the pure
Chern-Simons theory~\cite{jackiw,IK}.

In this paper, we address the problem of describing planar
noncommutative theories where $\theta$, the noncommutativity parameter, is
a space-dependent object. If the dependence of $\theta$ is sufficiently
smooth, this phenomenon can be studied within the deformation
quantization approach~\cite{ko1}, since it naturally allows for an
expansion in powers of $\theta$ and its derivatives.  We are, however,
interested in cases where $\theta$ is not necessarily smooth, namely, when
$\theta$ may have an appreciable variation over length scales of the order
of $\sqrt{\theta}$. For example, one may think of situations where $\theta$ has
first-order zeros in a certain region of the plane.

It is important to have the tools to describe that sort of situation,
since it may naturally occur in the Condensed Matter Physics context.
For example, when the relation between the magnetic field and the
effective mass is space-dependent; or one could want to study
interfaces that divide regions with different noncommutativity
parameters. If that interface is rather narrow, an expansion in powers
of $\theta$ and its derivatives will certainly be unreliable.

A way to deal with the case of a $\theta$ that depends on only one of the
variables has been presented in~\cite{CMS}.  Our approach is instead
based on the use of a particular mapping between the operatorial
representation of the theory, and its functional version. We construct
the noncommutative theory in a way that is in principle valid for a
more general $\theta$, although explicit results are presented for the case
of a $\theta$ that depends on one variable.

The analysis of theories with space-dependent noncommutativity has
been full of technical difficulties, both at the mathematical and
physical levels. Much effort has been devoted in recent years to
understand their fundamental properties.  In~\cite{cf1} Kontsevich's
construction is interpreted in terms of a path integral over a sigma
model. On the other hand, the relation between the noncommutativity
function and curved branes in curved backgrounds has been studied
in~\cite{cornalba}. Besides, in~\cite{das}, it is shown how to
construct $U(1)$ gauge-invariant actions when the noncommutativity is
space-dependent.  See also~\cite{mssw} for a formulation of gauge
theories on spaces where the commutator between space coordinates is
linear or quadratic in those coordinates.

The structure of this article is as follows: in
section~\ref{sec:operator} we set up the general framework, defining
the elements that are required to construct the noncommutative field
theory in the operatorial version of the algebra. In
section~\ref{sec:function}, we deal with the representation of the
theory in its functional form, namely, using functions with a
$\star$-product. These general results are applied, in
section~\ref{sec:x1only}, to the case in which $\theta ({\hat x})$ is an
invertible operator, and depends on ${\hat x}_1$ and ${\hat x}_2$ only
through a linear combination of them, i.e., $\theta({\hat x}_1, \hat x _2)=\theta(c_1
{\hat x}_1+c_2 {\hat x}_2)$.  In section~\ref{sec:boundary}, for a $\theta
({\hat x})$ with an analogous dependence, we allow for a null
eigenvalue, and discuss the physical consequences of that property.
Finally, in section~\ref{sec:concl}, we present our conclusions.

\section{Operatorial description}\label{sec:operator}
We shall consider Quantum Field Theories defined on a two-dimensional
noncommutative region generated by two elements, ${\hat x}_1$ and
${\hat x}_2$, which satisfy a local commutation relation
\begin{equation}
  \label{eq:gencommrel}
[ \hat x_j , \hat x_k ] \,=\,i \, \epsilon_{jk} \, \theta(\hat x)\;,
\end{equation}
where $j,k = 1,2$ and $\hat x_1$, $\hat x_2$ denote Hermitian
operators on a Hilbert space $\mathcal H$. $\theta({\hat x})$, also an
Hermitian operator, is a local function of ${\hat x}_1$ and ${\hat x}_2$.
The form of $\theta$ will be further restricted later on, when considering
some particular examples. That will allow us to derive more
explicit results, at least under simplifying assumptions.

We are interested in defining Field Theory actions for fields that
belong to the space ${\mathcal A}$, the algebra generated by $(\hat
x_1, \hat x_2)$.  To accomplish that goal, one has to introduce two
independent derivations (corresponding to the two coordinates $\hat x
_1$ and $\hat x _2$) plus an integration on ${\mathcal A}$.  Since the
fields, their products and linear combinations, are all elements of
${\mathcal A}$, one can, as usual, define the integral as the trace of
the corresponding product of fields.  Indeed,
\begin{equation}
  \label{eq:trace}
\int A(\hat x _i) \, \equiv \, Tr [ A(\hat x _i) ] \,,
\end{equation}
for any $A \in {\mathcal A}$, has the property of being linear and
invariant under cyclic permutation of the factors in a product.

Regarding the derivatives ${\hat D}_j$, in the present context they
are required to verify the following properties:
\begin{enumerate}
\item[(i)]the $\hat D_j$'s are linear operators;
\item[(ii)]they satisfy Leibniz' rule: ${\hat D}_j (A \, B)\,=\,( {\hat D}_j A
  ) \, B + A\, ( \hat D_j B)$;
\item[(iii)]the integral of a derivative vanishes: $Tr [\hat D_j A(\hat x)]
  =0$\,;
\item[(iv)]when $\theta \to \, {\rm const}$, $\hat D_j \to
  \partial_j$. This condition is not part of the formal
  definition of the derivatives, but we impose it in order to define
  actions that are comparable with their constant-$\theta$
  counterparts.
\end{enumerate}

Conditions (i)-(iii) are automatically satisfied if one uses {\em
  inner\/} derivations, i.e., those that can be written as
commutators: ${\hat D}_j A \, \equiv \, [d_j , A]$, with $d_j \in {\mathcal
  A}$.  When $\theta (\hat x)$ does have an inverse (denoted $\theta^{-1}(\hat
  x)$), in ${\mathcal A}$, a suitable choice for the $d_j's$ is given
  by the expression:
\begin{equation}
d_j \, \equiv \, \frac{i}{2} \,  \epsilon_{jk} \, \{  \theta^{-1}(\hat x) \,,\, \hat x_k \} \;,
\end{equation}
where $\{\,,\,\}$ denotes the anticommutator.
Conditions (i)-(iii) are then valid (as for any inner derivation);
regarding condition (iv), by acting on the generators of the algebra we see
that:
\begin{equation}
{\hat D}_j \hat x_k \,=\,\delta_{jk}+ \frac{i}{2}  \epsilon_{jl} \, \{ {\hat x}_l
\,,\, [  \theta^{-1}({\hat x}) \,,\, {\hat x}_k ]\} \;.
\end{equation}
Hence, condition (iv) is also fulfilled.

The final ingredient is the notion of adjoint conjugation. $A^\dag$ is
defined, as usual, by
\begin{equation}
 \label{eq:adjoint1}
 \langle f \vert A^\dag\vert g \rangle\,=\,\overline{\langle g \vert
   A\vert f \rangle } \;\;\; \forall \,
 f\,,\,g\;\; \in \mathcal H \;.
\end{equation}
Since $d_j^\dagger = - d_j$, we see that the derivative of an Hermitian
element of $\mathcal A$ will also be Hermitian.

We are now equipped to construct a noncommutative field theory in
$2+1$ dimensions, the simplest example being that of a scalar field
action $S$ for an Hermitian field $\phi$:
\begin{equation}
 \label{eq:operaction}
S\;=\;\int dt \, Tr \left[\frac{1}{2} D_\mu \phi (\hat x ,t) \,D_\mu \phi
  (\hat x ,t)+V(\phi) \right]
\end{equation}
where $D_\mu \equiv (\partial_t , \hat D_1, \hat D_2)$ (the time
coordinate is assumed to be commutative) and $V(\phi)$ is positive
definite.

Although this is, indeed, a perfectly valid representation for a
scalar field action on a noncommutative two-dimensional region, its
form is inconvenient if one has in mind its use in concrete (for
example, perturbative) calculations. Besides, the quantization of the
theory becomes problematic, and it is also rather difficult to compare
results with the ones of its commutative counterpart.

To address this problem, in the next section we consider the equivalent
description of the noncommutative theory in terms of functions equipped with a
noncommutative $\star$-product.

\section{Functional approach}\label{sec:function}
In this setting, the dynamical fields are not operators, but rather elements
of \mbox{$C^{\infty}(\mathbb R ^{2+1} )$}, and the (noncommutative) product
between the operators in ${\mathcal A}$ is mapped onto a noncommutative
\mbox{$\star$-product}.

This is, indeed, the idea behind the \textit{deformation
  quantization}~\cite{ca1}; the tools and ideas needed to deal with this
problem in a rather general setting have already been constructed (see, for
example~\cite{ca1} and~\cite{ko1}). Explicit expressions within this approach,
however, are difficult to derive, except when $\theta$ verifies certain
regularity conditions, which allow the resulting~\mbox{$\star$-product} to be
expressed by an expansion in powers of $\theta$ and its derivatives. By
`regularity conditions' we mean that $\theta$ has to be sufficiently smooth
for that expansion to converge. The measure of the smoothness is given
by the relation between those derivatives and the (only) other
dimensionful object, namely $\theta$. More explicitly, we should have:
$\partial_j \theta / \theta^{1/2} << 1$.
We want to consider here situations where this condition
for $\theta$ is not met (for example, when $\theta$ vanishes with a non-zero
derivative) so that an expansion in powers of $\theta$ and its
derivatives is either not possible or unreliable. We do that by using
a particular approach for the representation of the noncommutative
algebra of operators over the space of functions, which bypasses the
discussion on deformation quantization.  In this way, we shall obtain
a noncommutative $\star$-product which is valid even when such an
expansion makes no sense. This $\star$-product is based on the operatorial
approach of section~\ref{sec:operator}, and it follows from the
introduction of a specific mapping between operators and functions.

\subsection{Normal-ordering and kernel
  representations}\label{subsec:normalorder}
A one-to-one correspondence between operators $A(\hat x)$ and
functions $A(x)$ can be obtained by introducing a specific
operator-ordering prescription.  Here, we shall restrict the class of
operators considered to the ones that can be put into a
`normal-ordered' form. We define that form by the condition that all
the ${\hat x}_1$'s have to appear to the left of all the $\hat
x_2$ in the expansion of $A(\hat x)$ in powers of ${\hat x}_1$ and
${\hat x}_2$. Namely, we shall
consider the subspace ${\mathcal A}'$ of ${\mathcal A}$ that
consists of all the operators $A({\hat x})$ that can be represented as
follows:
\begin{equation}
A(\hat x) \;=\; \sum_{m,n} a_{mn} \; {\hat x}_1^m \, {\hat x}_2^n
\end{equation}
where the $a_{mn}$ are (complex) constants~\footnote{${\mathcal A}'$
  and ${\mathcal A}$ can coincide or be isomorphic. That would be the
  case, for example, if the Poincar\'e-Birkhoff-Witt property were
  satisfied for ${\mathcal A}$.}.

For some particular forms of $\theta$, any monomial in ${\hat x}_1$ and ${\hat
  x}_2$ can be converted into a normal-ordered form by performing a finite or
infinite number of transpositions.  For example, when $\theta$ is a
normal-ordered formal series (what we shall assume), we can map any
monomial in ${\hat x}_1$, ${\hat x}_2$ into normal order, albeit the
monomial will be now equivalent to an infinite normal-ordered series.
We shall later on consider a specific example which corresponds to a
much simpler situation: a $\theta$ which depends only on the variable
${\hat x}_1$. In that case, every monomial is equivalent to a
normal-ordered polynomial.

Thus for every operator $A({\hat x})$ in ${\mathcal A}'$, we have $A (\hat
x)\,=\, A_N (\hat x)\,$ where $A_N({\hat x})$ is its normal-ordered form.  We
assign to each $A({\hat x})$ the (c-number) function $A_N(x)$, obtained by
replacing in $A_N({\hat x})$ the operators $\hat x _i$ by commutative
coordinates $x _i$.  We then have a one-to-one mapping ${\mathcal S}$:
\begin{equation}
  \label{eq:mapnormal}
A(\hat x ) \, \to \, {\mathcal S}[A({\hat x})] \,=\, A_N (x) \;.
\end{equation}

To each operator $A({\hat x})$ we can also associate another function: its
`mixed' integral kernel $A_K(x_1,x_2)$
\begin{equation}
  \label{eq:mapkernel}
A_K (x_1, x_2)\,=\, \langle x_1\vert  A (\hat x _1, \hat x _2)\vert x _2\rangle \;,
\end{equation}
where $\hat x _i \, \vert x_i \rangle \,=\, x_i \,\vert x_i \rangle $.

The relation between $A_N$ and $A_K$ is quite simple; indeed:
\begin{equation}
  \label{eq:normal-kernel}
A_K (x_1, x_2)\,=\,  \langle x_1\vert x _2\rangle \, A_N (x_1, x_2 )\;.
\end{equation}
Since operator products are easily reformulated in terms of $A_K$, and we know
its relation to $A_N$, we use this relation to define the $\star$-product.

\subsection{Definition of the $\star$-product} \label{subsec:product}
To represent the algebra $\mathcal A$ on $C^{\infty}(\mathbb R ^{2+1})$, we
define the $\star$-product induced by the map (\ref{eq:mapnormal}):
\begin{equation}
  \label{eq:star1}
A_N \star B_N \,\equiv \, {\mathcal S}\left[ {\mathcal S}^{-1}(A_N) \,
{\mathcal S}^{-1}(B_N)\right]
\;.
\end{equation}
On the other hand, Equation (\ref{eq:normal-kernel}) may be used to obtain an
integral representation of the $\star$-product. Since:
\begin{equation}
(A \, B)_K (x_1, x_2)\,=\, \int d\tilde{x}_1 d\tilde{x}_2 \, A_K(x_1,
\tilde x _2)\, \langle \tilde x _2\vert \tilde x _1\rangle \, B_K(\tilde x _1
, x_2) \, ,
\end{equation}
we take (\ref{eq:normal-kernel}) and (\ref{eq:star1}) into account,
to arrive to the expression
\begin{equation}
  \label{eq:star2}
(A_N \star B_N) (x_1,x_2) = \int d\tilde{x}_1 d\tilde{x}_2 \,
\frac{\langle x _1\vert \tilde x _2 \rangle \langle\tilde x_2\vert
\tilde x_1 \rangle  \langle \tilde x_1\vert x _2\rangle}{\langle x _1
\vert x _2\rangle} \, A_N(x_1, \tilde x _2) \, B_N(\tilde x _1, x_2) \;.
\end{equation}
This product is evidently noncommutative, and it is also associative:
\begin{equation}
  \label{eq:associative}
(A_N \star B_N) \star C_N \, = \, A_N \star (B_N \star C_N) \;,
\end{equation}
a property which can be explicitly verified by using the definition of
the $\star$ product, or also by noting that the associativity of the
operator product is inherited by the $\star$ product (by an application of the
normal-order mapping).

Furthermore, it is immediate to prove that $(C^{\infty}(\mathbb R ^{2+1} ),
\star)$, henceforth noted as $C^{\infty}_\star$, reproduces the structure of
$\mathcal A$. Indeed, a straightforward calculation shows that:
\begin{equation}
  \label{eq:starcommutation}
[x_1 , x_2]_\star \, \equiv \, x_1 \star x_2 - x_2 \star x_1 \,=\, i \,
\theta_N (x) \;.
\end{equation}

Eq. (\ref{eq:star2}) is an integral representation of the noncommutative algebra (\ref{eq:gencommrel}), which depends on the function
\begin{equation}
F(x_1,x_2;\, \tilde x _1, \tilde x_2)\,=\,
\frac{\langle x _1\vert \tilde x _2 \rangle \langle\tilde x_2\vert
\tilde x_1 \rangle  \langle \tilde x_1\vert x _2\rangle}{\langle x _1
\vert x _2\rangle}\;.
\end{equation}
Constructing $F(x; \tilde x)$ for a general $\theta_N (x)$ is a very complicated problem; from Eq. (\ref{eq:starcommutation}) we may derive the integral equation
\begin{equation}
\int d\tilde x _1 d\tilde x _2 \, F(x_1,x_2;\tilde x _1, \tilde x _2)\,=\,x_1 x_2 - i \theta_N (x_1, x_2)\,.
\end{equation}
Even an expansion in powers of $\theta(x)$ is quite involved if $\theta$ has a general dependence on $(x_1,x_2)$. However, in section \ref{sec:x1only} we will show how to obtain explicit expressions for the simpler case $\theta=\theta(x_1)$.

\subsection{Integral, derivatives and adjoints in $C^{\infty}_\star$}~\label{subsec:intder}
 Based on the results of section~\ref{sec:operator} , we construct
the integral and derivatives now on $C^{\infty}_\star$. In what follows, to
simplify the notation, we suppress the `$N$' suffix when denoting a
normal symbol.

If $\theta(x)$ is everywhere different from zero, two possible inner
derivations are obtained by rewriting the ones of the operatorial
formulation, namely,
\begin{equation}
  \label{eq:functderiv2}
D_j A(x) \,\equiv \, [d_j(x) \, , \, A(x) ]_\star \;,
\end{equation}
where
\begin{equation}
  \label{eq:funcderiv}
d_j \, \equiv \, \frac{i}{2} \,  \epsilon_{jk} \, \{
\theta^{-1}( x) \,,\,  x_k \}_\star \;.
\end{equation}
In this equation $\theta ^{-1}$ is not the usual inverse function, but rather
the inverse w.r.t. the $\star$-product, i.e.,
$$
\theta ^{-1} (x) \equiv \mathcal{S} \big(\theta ^{-1} (\hat x) \big) \;.
$$

Since the $D_j$'s are $\star$-commutators, they are linear and obviously
satisfy the Leibnitz rule,
\begin{equation}
  \label{eq:leibniz2}
D_j (A \star B)\,=\, D_j A \star B + A \star D_j B \;,
\end{equation}
which is the $C^{\infty}_\star$ version of condition (ii) in
section~\ref{sec:operator}.  However, using the explicit expression
(\ref{eq:star2}) for the $\star$-products in the derivatives, we conclude
that
\begin{equation}
\int dx_1 dx_2 \, D_j (A(x)) \, \neq \, 0 \;,
\end{equation}
and
\begin{equation}
\int dx_1 dx_2 \, (A \star B)(x) \, \neq \, \int dx_1 dx_2 (B \star A)(x)
\, ,
\end{equation}
where $\int dx_1 dx_2$ is the usual integration over $\mathbb R^2$ with a
(flat) Euclidean metric. Hence,
neither integration by parts (with respect to $D_i$) nor cyclicity would be
valid if this definition of integral were used. Both of these properties,
which are essential in the construction of a sensible field theory, can
fortunately be satisfied if the factor $\vert \langle x _1\vert x _2\rangle
\vert ^2$ is included in the integration measure.  Thus, we define the
integral as:
\begin{equation}
  \label{eq:integral}
\int d \mu (x)\, A(x) \, \equiv \, \int dx_1 dx_2 \, \vert \langle x _1\vert x _2\rangle \vert^2 \, A(x) \;.
\end{equation}
The previous definition could also be derived from the operatorial trace, since
one notes that:
\begin{equation}\label{eq:integral2}
Tr \, A (\hat x) \,=\,\int dx_1 dx_2 \, \vert \langle x_1\vert
x_2\rangle \vert ^2 A_N(x)\;.
\end{equation}
Besides, the equalities
\begin{equation}
  \label{eq:parts}
\int d\mu (x) \, D_i (A(x)) \, = \, 0 \; ,
\end{equation}
\begin{equation}
  \label{eq:ciclic}
\int d\mu (x)\, (A \star B)(x) \, = \, \int d\mu (x)\, (B \star A)(x) \,
\end{equation}
are simple consequences of Eq. (\ref{eq:integral2}).

On the other hand, the adjoint defined in section~\ref{sec:operator}
can be represented in $C^{\infty}_\star$ by defining
\begin{equation}
  \label{eq:adjoint3}
A^\dag(x)\,=\,{\mathcal S} [A^\dag(\hat x)]\;.
\end{equation}
>From (\ref{eq:adjoint1}), (\ref{eq:mapkernel}) and
(\ref{eq:normal-kernel}), (\ref{eq:adjoint3}) can be represented
explicitly as
\begin{equation}
  \label{eq:adjoint4}
A^\dag(x_1,x_2)\,=\,\int d\tilde{x}_1 d\tilde{x}_2 \,
\frac{\langle x_1\vert \tilde x _2 \rangle \langle \tilde x_2\vert
\tilde x_1 \rangle  \langle\tilde x _1\vert x _2\rangle}{\langle x_1
\vert x _2\rangle} \,{\bar A} (\tilde  x _1 ,\tilde x _2)\;.
\end{equation}
While in the Weyl ordering prescription the hermiticity of operators is
tantamount to reality of functions, here $A^\dag(\hat x )=A(\hat x)$
translates into the condition:
\begin{equation}
  \label{eq:selfadj}
A (x_1,x_2)\,=\,\int d\tilde{x}_1 d\tilde{x}_2 \,
\frac{\langle x_1\vert \tilde x_2 \rangle \langle \tilde x_2\vert
\tilde x_1 \rangle  \langle \tilde x_1\vert x _2\rangle}{\langle x_1
\vert x _2\rangle} \,{\bar A} (\tilde  x _1 ,\tilde x _2)\;.
\end{equation}

\section{The case $\theta (x_1,x_2)=\theta(x_1)$}\label{sec:x1only}
It is evident that the knowledge of $\langle x_1| x_2\rangle$ plays a
fundamental role in the definition of the $\star$-product previously
introduced. In order to carry on a more detailed analysis, we restrict
ourselves here to a particular case, tailored such that $\langle x_1
\vert x_2 \rangle$ can be evaluated exactly. A simple way to
accomplish this is to consider the following form for $\theta (x)$:
\begin{equation}
  \label{eq:particular1}
[x_j,x_k]_\star \,=\,i \epsilon_{jk} \, \theta (c_1 x_1 + c_2 x_2 )\;.
\end{equation}
Under the redefinitions: $c_1 x_1 + c_2 x_2 \to x_1$, $x_2 \to x_2$, this
relation can be equivalently written as
\begin{equation}
  \label{eq:particular2}
[x_j,x_k]_\star \,=\,i \epsilon_{jk} \, \theta(x_1) \;.
\end{equation}

To obtain $\langle x_1 | x_2\rangle$, we come back to the
operatorial description of ${\hat x}_1$ and ${\hat x}_2$. The
spectra of those operators can be found by representing them on
the space of eigenfunctions of $\hat x_1$:
\begin{equation}
\hat x _1 |x_1 \rangle \,=\, x_1 \, | x_1 \rangle \;.
\end{equation}
On that space, the Hermitian operator ${\hat x}_2$ is represented by
\begin{equation}
{\hat x}_2 \,=\, -i \Big(\theta (x_1) \partial_1 + \frac{1}{2} \theta
'(x_1) \Big)  \;,
\end{equation}
where $\partial_1 \equiv \partial / \partial x_1$, $\theta '(x_1) \equiv \partial \theta / \partial x_1$.  To
find $\langle x_1 \vert x_2 \rangle$, we need the eigenvectors of ${\hat x}_2$ on the basis $\{\vert x _1 \rangle\}$.  Assuming that the operator $\theta({\hat x}_1)$ is invertible, which
is equivalent to saying that the function $\theta(x_1)$ has no
zeros~\footnote{This condition will be relaxed later on.}, we can
solve the corresponding differential equation to find:
\begin{equation}
  \label{eq:norm}
  \langle x_1 \vert x_2 \rangle \, = \, \frac{1}{\sqrt {2 \pi}} \, \theta^{-1/2} (x_1) \,
  {\exp} \Big(i x_2 \int^{x_1} \, dy_1 \, \theta^{-1} (y_1) \Big)\;,
\end{equation}
which has continuous normalization: $\langle x_2 | x_2'\rangle =
\delta(x_2-x_2')$.  The spectra of both operators ${\hat x}_1$ and ${\hat
  x}_2$ is the set of all the real numbers. This property is modified, as
we shall see, when the condition on the zeroes of $\theta$ is relaxed.

\subsection{Properties of the $\star$-product }\label{subsec:starprop}
Since $\vert \langle x_1 \vert x_2 \rangle \vert ^2 =\frac{1}{2
  \pi}\theta^{-1}(x_1)$, the integration measure in (\ref{eq:integral})
becomes:
\begin{equation}
  \label{eq:mumeasure}
d\mu (x) \, \equiv \, \frac{1}{2 \pi}\, dx_1 dx_2 \, \theta^{-1}(x_1)\,.
\end{equation}
The noncommutative product of (\ref{eq:star2}) reduces to
\begin{equation}
  \label{eq:star3}
  (A \star B)(x) \,=\, \int d\mu (\tilde x) \, {\exp} [ i (x_2 - \tilde x _2)
  \, \Delta g (x_1, \tilde x _1 ) ] \, A(x_1, \tilde x _2) \, B(\tilde x _1
  , x_2) \,,
\end{equation}
where
\begin{equation}
g(x_1)\equiv \int ^ {x_1} dy_1 \theta^{-1}(y_1) \;\;, \;\; \Delta g (x_1,
\tilde x _1) \equiv g(\tilde x _1) - g(x_1) \;.
\end{equation}

By using some elementary algebra, we derive the useful relations:
\begin{equation}
  \label{eq:rel1}
A(x_1) \star B(x_1,x_2)\,=\,A(x_1) B(x_1,x_2)
\end{equation}
\begin{equation}
  \label{eq:rel2}
A(x_1,x_2) \star B(x_2)\,=\,A(x_1,x_2) B(x_2)
\end{equation}
\begin{equation}
  \label{eq:rel3}
A(x_2) \star B(x_1)\,=\,A\Big(-i \theta(x_1) \partial_1+x_2\Big) B(x_1)\;.
\end{equation}

Furthermore, (\ref{eq:rel1})-(\ref{eq:rel3}) may be used to obtain an
alternative expression for the $\star$-product. Writing the generic
normal-ordered function as
\begin{equation}
  \label{eq:normalfunction}
A(x_1,x_2)= \sum_n \, \alpha^A_n (x_1)\,\beta^A_n (x_2) \;,
\end{equation}
we see that
\begin{equation}
  \label{eq:star4}
  A(x_1,x_2)\star B(x_1,x_2)\, = \, \sum_{n,m}\,\alpha^A_n (x_1)\,\beta^A_n
  \Big(-i \theta(x_1) \partial_1+x_2\Big)\alpha^B_m (x_1)\,\beta^B_m (x_2)\;.
\end{equation}
Either the form (\ref{eq:star3}) or (\ref{eq:star4}) may prove to
be more useful, depending on the context. For instance, if an
expansion in powers of $\theta (x_1)$ and its derivatives is
valid, Eq. (\ref{eq:star4}) gives
 $$
A_N(x) \star B_N (x)\,=\,A_N (x) B_N (x) - i \theta(x_1) \;
\partial_2 A_N (x) \, \partial_1 B_N (x) +
$$
\begin{equation}
   \label{eq:star5}
 - \frac{1}{2}\,\theta^2(x_1) \;\partial_2 ^2 A_N (x) \, \partial_1 ^2  B_N (x)- \frac{1}{2}\,\theta^2(x_1)\theta'(x_1) \;\partial_2 ^2 A_N (x) \, \partial_1   B_N (x) \, + \, \cdots \;.
 \end{equation}

Let us conclude by considering the derivatives for the present
case. From the operatorial construction we have the general
expression:
\begin{equation}
D_j A(x)\,=\, [ d_j(x) , A(x)]_\star \;,
\end{equation}
where
\begin{equation}
d_j(x) \,=\, \frac{i}{2} \, \epsilon_{jk} \, \{ \theta^{-1}(x) \,,\, x_k \}_\star \;.
\end{equation}
Applying relations (\ref{eq:rel1})-(\ref{eq:rel3}), we may simplify
the expressions for the $d_j$'s for the particular case $\theta(x) =
\theta(x_1)$:
\begin{eqnarray}
d_1 (x) &=& i \, \theta^{-1} (x_1) \, x_2 \, -\, \frac{1}{2} \, \theta^{-1}(x_1) \partial_1
\theta(x_1) \nonumber\\
d_2 (x) &=& - i \, \theta^{-1}(x_1) \, x_1 \;.
\end{eqnarray}
Then the action of the $D_1$ derivative on $A(x_1,x_2)$ may be
written as follows:
$$
D_1 A(x) \;=\; \partial_1 A(x) \,+\, i x_2 \, [\theta^{-1}(x_1) \,,\, A(x)
]_\star
$$
\begin{equation}\label{eq:derivx}
\,-\,\frac{1}{2} \, [ \theta^{-1} (x_1) \partial_1 \theta(x_1) \,,\,  A(x) ]_\star\;.
\end{equation}
It is worth noting that, since $\theta$ depends only on $x_1$, $\theta
^{-1}(x_1)$ coincides with the usual inverse function

On the other hand, $D_2$ is given by:
\begin{equation}
D_2 A(x)\,=\,i [\theta^{-1}(x_1) x_1 , \,A(x)]_\star \;,
\end{equation}
and it does not lead immediately to a similarly simple expression,
involving a detached term with $\partial_2$. Indeed, after a little algebra one sees that
\begin{equation}
D_2 A(x)\,=\, \Big( 1 - x_1 \theta^{-1}(x_1) \partial_1 \theta(x_1) \Big) \partial_2 A(x) \,+\, \ldots
\end{equation}
where the omitted terms involve higher powers of $\partial_2$ acting on $A$.

However, the problem of coping with the previous expression for $D_2$ can
be entirely avoided by recalling that, since the noncommutative function $\theta(x)$ depends only on $x_1$, a simpler definition of a derivative should exist as a reflection of the invariance of $\theta$ under $x_2$ translations.
 Indeed, the outer derivative
\begin{equation}
  \label{eq:derivy}
D_2 A(x)\,\equiv\,\partial_2 A(x)\;,
\end{equation}
satisfies all the properties of a derivation:
\begin{equation}
\partial_2 (A \star B)\,=\, \partial_2 A \star B + A \star \partial_2 B
\;\;, \;\; \int d\mu (x) \partial_2 A(x)=0
\end{equation}
for any $A(x)$ vanishing at infinity. We shall henceforth
assume that $D_2$ stands for (\ref{eq:derivy}), while $D_1$
corresponds to (\ref{eq:derivx}).

Of course, $D_1$ and $D_2$ cannot be simultaneously diagonalized (a
property that also holds true when both are {\em inner\/}
derivatives).  Their commutator reads
\begin{equation}
  \label{eq:commderiv}
[D_1,\,D_2]A(x)\,=\,-i  [\theta^{-1}(x_1) , \,A(x)]_\star \;,
\end{equation}
which is akin to a noncommutative curvature. The relation between Poisson structure and curvature has been considered
in~\cite{madore2}.

This completes our discussion on the tools required to construct a
field theory over $C^\infty_\star$. Again, the simplest case corresponds to the
noncommutative generalization of a real scalar field action $S(\phi)$:
\begin{equation}
S(\phi) \, = \, l^2 \,\int dt \,d\mu(x) \,\Big( \frac{1}{2} D_\nu \phi(x) \star D_\nu
\phi(x)+V_\star(\phi)\Big)\;,
\end{equation}
which is the functional transcription of the operatorial action
(\ref{eq:operaction}). We have introduced the parameter $l$, with the
dimensions of a length, in order to have a dimensionless action $S$.
$l$ can be naturally associated to the typical length defined by
$\sqrt{\theta (x_1)}$. $\phi$ satisfies the constraint (\ref{eq:selfadj}) and
$V_\star[\phi(x)]=\mathcal S \Big(V[\phi(\hat x)] \Big)$ is assumed to be
positive.

\subsection{Interpretation}
The formulae (\ref{eq:derivx}), (\ref{eq:derivy})
and (\ref{eq:commderiv}) have an intuitive physical interpretation in
terms of the Landau problem \cite{landau}. For a charged particle of
mass $m$ moving in the plane in the presence of a perpendicular
magnetic field depending only on one of the coordinates: $B=B(x_1)$,
the Lagrangian is
$$
L\,=\,\frac{1}{2} \, m \,\dot{ x}_i^2 -\dot{x}_i  A_i (x) \;.
$$
Since $B=\partial_1 A_2 - \partial_2 A_1$, we can choose, for the vector potential,
$$
A_1\,=\,-x_2\,B(x_1)+\varphi(x_1) \;\;\;\;,\;\;\; A_2\,=\,0 \;,
$$
where $\varphi(x_1)$ accounts for the remanent gauge freedom. The mechanical momentum operators, defined as
$$
\hat \pi _j \,=\,-i \partial_j+A_j(\hat x )\;,
$$
have the commutation relations
$$
[\hat \pi _1,\hat \pi_2]\,=\,-i\,B(\hat x _1) \;.
$$
Their utility comes from the fact that the Hamiltonian is
$$
H\,=\,\frac{1}{2m} \pi_i ^2 \;.
$$

Remembering that the noncommutativity (\ref{eq:particular2}) is
associated to the reduction to the lowest Landau
level~\cite{flhigh,jackiw} , we are naturally led to identify the
$D_j$ with the mechanical momenta: $D_j \to i \hat \pi _j$. Then $
-[\theta^{-1}(x_1)\, ,\, ]_\star$ is interpreted as the noncommutative
generalization of a magnetic field and the $D_j$ correspond to the
gauge choice
$$
A_1\,=\, x_2 \, [\theta^{-1}(x_1) \,,\,]_\star
\,+\,\frac{i}{2} \, [ \theta^{-1} (x_1) \partial_1 \theta(x_1) \,,\,  ]_\star\;\;\;\;,\;\;A_2\,=\,0 \;.
$$

\section{Boundary contribution}\label{sec:boundary}
In this section we study the consequences of extending the previous
formulae to a case in which $\theta(x_1)$ has a zero. As we will see,
this has interesting physical consequences, which we shall study for
the case of a theory defined by a noncommutative scalar field action.

Assuming that $\theta(x_1)$ has only one zero, at $x_1=0$, $D_1$ and $\int d\mu
(x)$ are ill-defined at $x_1=0$; furthermore, from the eigenvalue
equation for $\langle x_1 \vert x_2\rangle $, one finds that $\langle x_1 \vert x_2 \rangle$
exists and it is unique only for $x_1 \, \in \, (0,\infty)$ or $x_1 \, \in \,
(-\infty,0)$, but not for the whole real axis. A restriction of the
operators to only one of those intervals naturally suggests itself.
We shall represent the operators $\hat x_1$ and $\hat x_2$ over the
subspace corresponding to the eigenvalues in the interval $(0,\infty)$.
Note that the presence of a zero in $\theta$ has led naturally to the
existence of a boundary: $(x_1=0,x_2)$ in the configuration space.
This boundary corresponds to the region where the coordinates commute,
and it defines a (lower-dimensional) commutative theory.

To deal with the singularities at $x_1=0$, we introduce a parameter
$\varepsilon$ such that $x_1 \, \in \, (\varepsilon,\infty)$, and $\varepsilon \, \to \, 0$ amounts to
approaching the boundary (which cannot be exactly reached, since some
operations would be ill-defined there). At the operatorial level, this
restriction can be achieved by the introduction of a projection
operator $P_\varepsilon (\hat x _1)$, such that $P_\varepsilon (x_1)=H(x_1-\varepsilon)$, where
$H(x_1-\varepsilon)$ is  Heaviside's step function. For instance, the trace over $x_1
\, \in \, (\varepsilon,\infty)$ can be `regulated' (to avoid the boundary) as follows:
\begin{equation}
Tr \Big(P_\varepsilon (\hat x _1) A(\hat x) \Big) \;.
\end{equation}
Translating this into the functional language, this means to consider
regulated integrals $I_{\varepsilon}$, of the form:
\begin{equation}
I_\varepsilon\,=\, \int d\mu(x) \, P_\varepsilon (x_1) \star A(x) \;,
\end{equation}
with $d\mu(x)$ given by (\ref{eq:mumeasure}). Recalling (\ref{eq:rel1}),
this integral can always be written as:
\begin{equation}
I_\varepsilon \,=\, \int d\mu(x) \, P_\varepsilon (x_1) \, A(x) \;.
\end{equation}
On the other hand, the derivatives $D_i$ defined in Eqs.
(\ref{eq:derivx}) and (\ref{eq:derivy}) are well-defined on $x_1 \, \in
\, (\varepsilon,\infty)$.  However, due to the presence of the projector in the
integration measure, the {\em regulated\/} integral of a derivative is
no longer zero:
\begin{equation}
\int d\mu (x)\,  P_\varepsilon (x_1) \, D_i A(x)\,=\,-\int d\mu (x)\, \Big(D_i  P_\varepsilon (x_1) \Big) \,A(x) \;.
\end{equation}
Since, from (\ref{eq:derivx}),
$$
D_i P_\varepsilon (x_1)\,=\,\delta_{1i} \, \partial_1 P_\varepsilon (x_1)
\,=\,\delta_{1i} \, \delta(x_1 - \varepsilon) \;,
$$
we arrive to
\begin{equation}
  \label{eq:bound1}
\int d\mu (x)\,  P_\varepsilon (x_1) \, D_i A(x)\,=\,-\delta_{1i}\, \int \frac{dx_2}{2 \pi \, \theta(\varepsilon)} \, A(x_1 =\varepsilon, x_2) \;.
\end{equation}
Therefore, a boundary contribution is generated. This property is to
be expected from the physical point of view, since there should be a
positive `jump' in the number of degrees of freedom when the theory
becomes commutative.

Let us apply the previous procedure to a scalar field,  whose
regulated action is:
\begin{equation}
  \label{eq:boundaction}
S\,=\, \frac{l^2}{2} \int dt \, d\mu (x) \,  P_\varepsilon (x_1) \, D_\nu \phi(x) \star D_\nu \phi(x) \;,
\end{equation}
where $D_\nu=(\partial_0,D_1,\partial_2)$. After integrating by parts, and applying
(\ref{eq:bound1}), we see that:
\begin{equation}
S\;=\;\frac{l^2}{2}\, \int dt \, d\mu (x) \,  P_\varepsilon (x_1) \, \frac{1}{2}\big(
\phi(x) \star D^2 \phi(x)+ D^2 \phi(x) \star  \phi(x)\big) \,+\, S_b
\end{equation}
where:
\begin{equation}
 \label{eq:boundaction2}
S_b \;\equiv \; -\frac{l^2}{4 \pi \, \theta(\varepsilon)} \, \int dt \, dx_2\;
\frac{1}{2}\big(\phi(x) \star D_1 \phi(x)+ D_1 \phi(x) \star \phi(x) \big) \Big \vert _{x_1=\varepsilon}\;.
\end{equation}
The first (`bulk') term does not generate any boundary contribution. However,
the second term $S_b$, which is a by-product of the zero at $x_1=0$, is
a boundary term.

$S_b$ is a $1+1$-dimensional action on the boundary $x_1=\varepsilon$
which, in general, has a complicated dynamics. To derive a more explicit
form, we take into account (\ref{eq:star4}) and (\ref{eq:derivx}) to
write:
$$
D_1 \phi(x)\,=\,\partial_1 \phi (x)+i x_2 \, \theta^{-1}(x_1) \, \phi(x) -i x_2 \sum_r \, a_r (x_1) \, \times
$$
$$
\times \; b_r\big(-i\theta(x_1)\partial_1+x_2\big)\,\theta^{-1} (x_1) \,-\, \frac{1}{2}
\partial_1\ln \theta(x_1) \; \phi (x)
$$
\begin{equation}
  \label{eq:bounderiv1}
+ \frac{1}{2} \sum_r \, a_r (x_1)\, b_r\big(-i\theta(x_1)\partial_1+x_2\big)\; \partial_1\ln
  \theta(x_1) \;,
\end{equation}
where we used the expansion:
\begin{equation}
\phi(x_1,x_2)\,=\,\sum_r  a_r (x_1) b_r (x_2) \;.
\end{equation}
Besides, we have
\begin{equation}
  \label{eq:bounderiv3}
  \phi(x) \star D_1 \phi (x)\,=\,\int d\mu (\tilde x) \, P_\varepsilon (\tilde x_1 )
  \,e^{i(x_2-\tilde x _2)\Delta g(x_1,\tilde x _1)}\, \phi(x_1,\tilde x _2) \,
  D_1 \phi(\tilde x _1 , x_2) \;,
\end{equation}
and an analogous expression for  $\phi(x) \star D_1 \phi (x)$

A non-trivial boundary contribution is then derived, whose explicit
behaviour depends upon the precise form of $\theta (x_1)$. Note that the
$\star$-product, when conveniently expanded, will introduce also
derivatives with respect to $x_2$. This is illustrated by the
following example.

\subsection{Example}\label{subsec:example}
The simplest situation occurs when
\begin{equation}
  \label{eq:thetaex}
\theta (x_1)\,=\,l g \,x_1 \,,
\end{equation}
where the dimensionless parameter $g$ controls the `strength' of the
noncommutativity. Here we shall perform an expansion valid for $g \ll
1$, as this approximation serves to the purpose of exhibiting a local
form for the term (\ref{eq:boundaction2}).  For an arbitrary $g$, the
term is of course non local.

The first step is to compute
\begin{equation}
  \label{eq:gex}
\Delta g (x_1,\tilde x _1)\,=\,(l g)^{-1}\, \ln \frac{\tilde x _1}{x_1}\;.
\end{equation}
Replacing (\ref{eq:thetaex}) into (\ref{eq:bounderiv1}), we have
\begin{equation}
  \label{eq:derivex}
D_1 \phi(x_1,x_2)\,=\,\partial_1 \phi (x_1,x_2)+\frac{x_2 +
    i l g / 2}{x_1}\,
\frac{\phi(x_1,x_2+il g)-\phi(x_1,x_2)}{i l g} \;.
\end{equation}
When $g$ is very small (at $l$ fixed),
$$
D_1 \phi(x_1,x_2)\,= \,\partial_1 \phi (x_1,x_2)+\frac{x_2}{x_1} \partial_2 \phi(x_1,x_2)
$$
\begin{equation}
  \label{eq:derivex1}
+ \frac{i l g}{2} \,\left[ \frac{x_2}{x_1}  \partial^2_2\phi(x_1,x_2) \,+\,
\frac{1}{x_1} \, \partial_2 \phi(x_1,x_2) \right] \,+\, \ldots \;.
\end{equation}
In $\phi \star D_1 \phi+D_1 \phi \star \phi$ there is a factor
$$
e^{i(x_2-\tilde x _2)\Delta g(x_1,\tilde x _1)}\,=\,e^{i(l
g)^{-1}(x_2-\tilde x _2) \ln(\tilde x _1 / x_1)} \;.
$$
so in the $g \to 0$ limit the stationary phase approximation is
valid. To express $\phi \star D_1 \phi+D_1 \phi \star \phi$ as a local series in
powers of $g$, we have to expand around
$x_2$: $\tilde x_2\,=\,x_2+g\, \xi $. After some algebraic manipulations,
we arrive to the leading contribution for $\varepsilon \to 0$:
\begin{equation}\label{eq:starex}
S_b \,=\, \frac{1}{2}\,\frac{l}{4 \pi g \varepsilon^2}\, \int dt \, dx_2 \,\left[  \phi^2
+(g l)^2 ( \partial_2\phi)^2 -\frac{2}{3}(g l)^2 \, x_2 \, \phi \, \partial_2^3 \phi +
O(g^2 \varepsilon, g \varepsilon ^2) \right] \;,
\end{equation}
where the field is evaluated at $x_1=\varepsilon$.  To get a reduced field with
the proper ($1+1$-dimensional) canonical dimension, we make the
redefinition:
\begin{equation}
{\tilde \phi}(x_2) \;=\; (\frac{1}{4\pi})^{1/2} \, \frac{l}{\varepsilon} \, \sqrt{g
l} \, \phi(\epsilon,x_2) \;.
\end{equation}
Hence:
\begin{equation}\label{eq:actionex}
S_b \,=\, \frac{1}{2}\, \int dt \,dx_2 \,\left[ (\partial_2{\tilde \phi})^2 \,+\,
\frac{1}{(g l)^2} {\tilde \phi}^2 -\frac{2}{3} x_2 \, {\tilde \phi}\,
\partial_2^3 {\tilde \phi} + \ldots \right] \;.
\end{equation}

Since the reality condition (\ref{eq:selfadj}) for $\theta$ given by Eq. (\ref{eq:thetaex}) can be written as
$$
\phi(x)\,=\,\bar{\phi}(x)+O(g)\,,
$$
$S_b$ contains a real (up to order $g$) mass term contribution in 1+1
dimensions, with a mass $M$ inversely proportional to $g$, given by
\begin{equation}
  \label{eq:massex}
M\,=\,\frac{1}{g l}\;.
\end{equation}

The action is not translation invariant, something that can be
understood as a relic of the existence of an external (non constant)
magnetic field. Furthermore, it is time-independent, and it may be
interpreted as proportional to the static energy of the boundary.

Note that this action comes from the first two terms in a small-$g$
expansion for $\phi \star D_1 \phi +D_1 \phi \star \phi$, and it already has some physical
information. The leading contribution, for example, goes like
$g^{-2}$, and is precisely the kind of term that one would introduce
to enforce the Dirichlet boundary condition $\phi(x_1=0,\,x_2)=0$. The
higher-order corrections (also at order $1/\varepsilon$) yield terms containing
derivatives of the scalar field on the boundary.

\section{Conclusions}\label{sec:concl}
In this paper we have discussed some aspects of space-dependent planar
noncommutativity, based on a particular mapping between normal-ordered
operators and functions. The use of such mapping has proved to be
quite useful, since it allows one to derive explicit forms for the
basic tools of the corresponding noncommutative field theory.

Guided by the operatorial description of section~\ref{sec:operator},
in section~\ref{sec:function} we defined an integral, derivatives, and
adjointness on $C^{\infty}_\star$.  A further restriction to the case
$\theta(x_1,x_2)=\theta(x_1)$ allowed us to compute explicitly the $\star$-product
obtaining the rather simple expression~(\ref{eq:star3}) and other
useful relations~(\ref{eq:rel1})-(\ref{eq:star4}). Results for that
case are consistent with the ones of~\cite{CMS} (equations (20)-(28)
of that paper), in the sense that the square root of the metric is
related to $\theta$ in the same way we have found to be the case in the
integration measure.

Equation~(\ref{eq:star3}) is valid for quite general functions
$\theta(x_1)$.  In particular, in section~\ref{sec:boundary}, we showed how
to generalize our approach to a function $\theta (x)$ vanishing along the
line $x_1=0$. This is a very interesting situation, since in the line
$x_1=0$ there is a transition from a noncommutative theory to a
commutative one. Of course, the transition is not continuous and many
objects from the noncommutative theory are ill-defined over that
region. The same situation appears when one takes the limit $\theta \to 0$ in
the constant-$\theta$ case. Therefore, the zero creates a boundary in the
configuration space that cannot be reached. We proved, starting from a
noncommutative scalar field theory, that there are boundary
contributions to the action, deriving the explicit form of its first
few terms for the case $\theta (x_1)=\alpha x_1$.

A natural question that arises at this point refers to the
relation between this approach to construct a noncommutative
$\star$-product and deformation quantization results, as stated in
\cite{ko1}. If an expansion in powers of $\theta (x_1)$ and its
derivatives is valid, then the $\star$-product defines a
deformation quantization. Indeed, Eq. (\ref{eq:star5}) defines,
according to ~\cite{ko1,ca1}, a star product.


\section*{Aknowledgments} G.~T.\ is supported by Fundaci\'on Antorchas,
Argentina.  C.~D.~F.\ is supported by CONICET (Argentina), and by a
Fundaci\'on Antorchas grant. G.~T. gratefully thanks A. Cabrera for
useful comments and discussions.  We thank F.~A.~Schaposnik for interesting
comments and explanations.


\begin{thebibliography}{99}
\bibitem{DN}M.~Douglas and N.~Nekrasov, Rev.~Mod.~Phys.~73: 977-1029 (2001).
\bibitem{szabo} R. ~ Szabo. Phys. Rept. {\bf 378}, 207 (2003).
\bibitem{castellani}L.~Castellani, Class.~Quant.~Grav.~17, 3377 (2000).
\bibitem{susskind}L.~Susskind, hep-th/0101029.
\bibitem{PP}A.~P.~Polychronakos, JHEP {\bf 0104} 011 (2001).
\bibitem{KS}D.~Karabali and B.~Sakita, Phys.~Rev.~{\bf B64}, 245316
(2001).
\bibitem{PA}V.~Pasquier, Phys.~Lett.~{\bf B 490}, 258 (2000).
\bibitem{LMR}B.~H.~Lee, K.~Moon and C.~Rim, Phys.~Rev.~D {\bf 64}, 085014
(2001).
\bibitem{landau} L. Landau, E. Lifshitz, {\textit {Quantum Mechanics}}, vol. 3, Butterworth-Heinemann, 3rd edition, Oxford, 1977.
\bibitem{gj}S.~M.~Girvin, and T.~Jach, Phys.~Rev. {\bf B 29}, 5617 (1984).
\bibitem{flc} M. ~Ciccolini, C.D. ~Fosco and A. ~Lopez. J.Phys.A {\bf 35},
10077 (2002).
\bibitem{jackiw}G.~V.~Dunne, R.~Jackiw, and C.~A.~Trugenberger.  Phys.
  Rev {\bf D41}, 661 (1990).
\bibitem{IK}I.~I.~Kogan, Int.~J.~Mod.~Phys.~A9, 3887 (1994).
\bibitem{ko1}M.~Kontsevich, ``Deformation quantization of
  Poisson manifolds, I,'' q-alg/9709040.
\bibitem{CMS} D.~H.~Correa, E.~F.~Moreno and F.~A.~Schaposnik. Phys.\ Lett.\ B {\bf 543}, 235 (2002).
\bibitem{cf1} A.~Cattaneo and G.~ Felder. Commun.Math.Phys
. {\bf 212}, 591 (2000).
\bibitem{cornalba} L.~Cornalba and R.~Schiappa. Commun. Math. Phys. {\bf 225},
33 (2002).
\bibitem{das} A.~Das and J.~Frenkel. Phys. Rev. D {\bf 69}, 065017 (2004).
\bibitem{mssw} J.~Madore, S.~Schraml, P.~ Schupp and J.~Wess. Eur. Phys. J. C
{\bf 16}, 161 (2000).
\bibitem{ca1} A. ~Cattaneo,``Formality and star products'', math.qa/0403135.
\bibitem{madore2} J.~Madore. Rept. Math. Phys. {\bf 43}, 231 (1999).
\bibitem{flhigh}C.~D.~Fosco, A.~Lopez. J.~Phys.~{\bf A37}, 4123 (2004).
\end{thebibliography}
\end{document}